Revised version for *Weather*

# Lightning detection in planetary atmospheres

Karen L. Aplin[1] and Georg Fischer[2]

1. Department of Physics, University of Oxford, Denys Wilkinson Building, Keble Road, Oxford OX1 3RH UK
2. Space Research Institute, Austrian Academy of Sciences, Schmiedlstr. 6, A-8042 Graz, Austria

## Abstract

Lightning in planetary atmospheres is now a well-established concept. Here we discuss the available detection techniques for, and observations of, planetary lightning by spacecraft, planetary landers and, increasingly, sophisticated terrestrial radio telescopes. Future space missions carrying lightning-related instrumentation are also summarised, specifically the European ExoMars mission and Japanese Akatsuki mission to Venus, which could both yield lightning observations in 2016.

## 1. Introduction

Lightning outside Earth's atmosphere was first detected at Jupiter by the Voyager 1 spacecraft in March 1979 (Smith et al., 1979; Gurnett et al., 1979). Since then, lightning has been detected on several other planets (e.g. Harrison et al, 2008), and it may even exist outside our solar system (e.g. Aplin, 2013; Hodosán et al., 2016). Beyond simple scientific curiosity, there are several reasons to study planetary lightning. The famous Miller and Urey experiment of the 1950s found that electrical discharges generated in conditions mimicking the early Earth's atmosphere produced amino acids, the starting point for life (e.g. Parker et al, 2011). More recent work has shown that energetic electrons accelerated in the electric fields in Martian dust storms can affect atmospheric chemistry, which may also be relevant for life (Harrison et al., 2016). The possibility of life elsewhere in the universe, and the conditions needed for it, remains one of the biggest scientific questions facing mankind, and provides a major motivation.

A further motivation is that the detection of lightning can be used to deduce more information about a planetary environment. Firstly, all known lightning generation mechanisms (discussed later in Section 4) involve atmospheric convection, so the presence of lightning can be an indicator of convection. Secondly, the low-frequency Schumann resonance signals established by lightning between the conductive surface and upper atmosphere can potentially be used to deduce the presence of water and volatiles in the envelopes of gas planets. This understanding can contribute to fundamental questions about the evolution of the solar system and the origins of life (Simões et al., 2012).

Lightning can also be used to deduce properties of the propagating medium, e.g. peak electron densities of a planet's ionosphere can be retrieved from the cut-off frequency of lightning radio emissions. Lightning can drive a global electric circuit, which leads to the presence of an electric field in the atmosphere away from storm areas; this may have subtle effects on particle transport and other atmospheric processes (e.g. Aplin, 2013). On a more practical note, the hazards from lightning may be relevant for *in situ* planetary instrumentation, and human exploration (e.g. Lorenz, 2008).

In this article we discuss the signals produced by lightning and how they can be exploited for detection. We then briefly review existing planetary lightning observations, focusing on recent findings, and conclude by looking to the future.

## 2. Signatures of lightning

*2.1 Optical emission*

Lightning produces optical emission from intense heating of the channel it makes in the atmosphere. Planetary lightning can be photographed at night with a simple, multipurpose camera, not specifically designed for lightning detection (Smith et al, 1977), as can be seen in Figure 1, which was the first image of extra-terrestrial lightning, on Jupiter. This photograph, from 1979, is a remarkable achievement considering that reports of terrestrial lightning detection from space start in 1978 (Labrador, 2016).

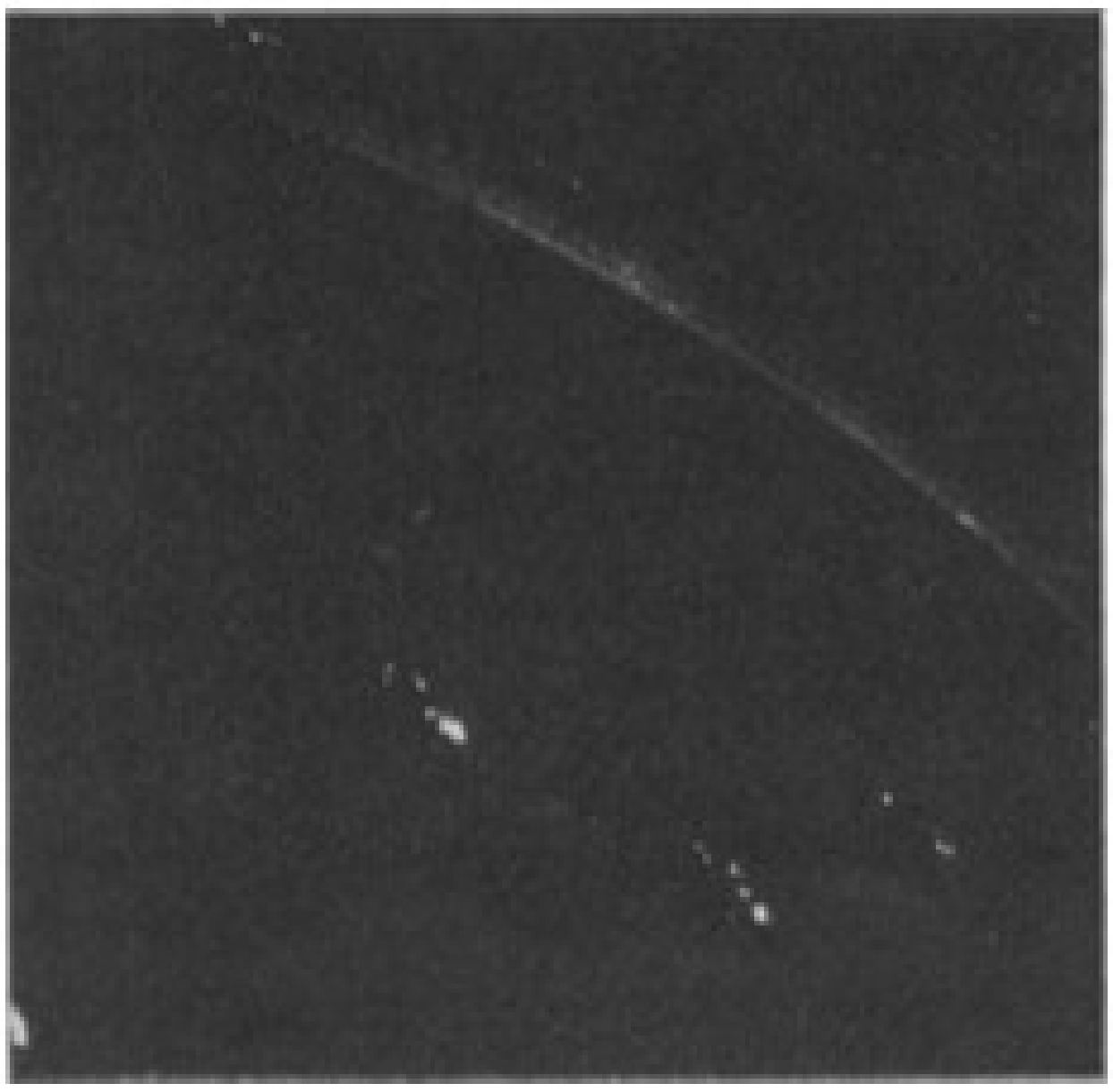

**Figure 1. Lightning detected on the night side of Jupiter by Voyager 1. The view looks towards the north pole, and aurorae can also be seen. The visible arc of limb is about 30 000 km long. The flashes have an optical energy of about $10^9$J (Borucki et al., 1982). Reproduced with permission from Smith et al. (1979).**

Jovian lightning was readily detectable by Voyager 1 due to its substantial optical energy ($10^9$J) and spatial scale, and it has been photographed by every subsequent Jupiter flyby. However, more sensitive instrumentation is usually needed for optical detection of lightning. Cameras with specialised filters "tuned" to the spectral emission peaks expected for the atmosphere of interest are often used. These lines were first identified in laboratory experiments (Borucki et al., 1985), which found that the terrestrial lightning spectrum is strongest in the infrared nitrogen absorption lines. Aircraft measurements subsequently indicated a stronger signal at 777.4 nm from an oxygen excitation line, which was ultimately chosen for satellite instruments (Christian et al., 1989). Use of such filters permits lightning detection during the day. The 777.4 nm excitation line is also predicted from Venusian lightning in its carbon dioxide atmosphere. Jupiter, Saturn, Uranus and Neptune are all expected to emit in the red at 656 nm because of their hydrogen atmospheres, but the spectral signature of lightning from the giant planets is still poorly understood. At Saturn optical flashes were first detected on the night side with a broadband visible filter (Dyudina et al., 2010), and on the dayside with a blue filter (Dyudina et al., 2013), whereas on Jupiter, nightside emissions at 656 nm were ten times weaker than expected (Dyudina et al., 2004). Transient luminous events (TLEs) occurring above thunderstorms immediately after lightning like sprites, elves or jets have been thoroughly investigated on Earth in the last two decades. So far no TLEs have been observed on any planet other than Earth, but their existence, and prospects for detection through their optical properties, were considered by Yair et al. (2009).

*2.2 Radio emission*

As well as producing optical signals, the ionised channel acts as an antenna for broadband radio emissions, often called "sferics". The frequencies emitted depend on the stroke duration and time profile; small-scale discharges (such as "glow" or corona discharge) emit at a higher frequency than longer channels. Detection of these signals from space is limited by a combination of the energy of the strike and its frequency spectrum. Sferics can be detected outside an atmosphere if they exceed its ionospheric cut-off frequency. For most planets this cut-off is a few MHz, therefore we define high frequency (HF) emissions as those that can escape from the planet into free space, about 3-30 MHz. The ionospheric cut-off frequency is about 10 MHz for Earth, and provides a lower limit for detection of extra-terrestrial lightning by terrestrial radio telescopes (Zarka et al., 2004).

Whistlers are very low frequency (VLF; about 20 Hz-20 kHz) electromagnetic waves generated by lightning, which are guided along magnetic field lines and therefore detectable by spacecraft in planetary magnetospheres. The ultra low frequency (ULF) "Schumann resonances" of a few Hz indirectly generated by lightning are only detectable *in situ* because their low frequency prevents the signals from propagating outside the ionosphere, but until now, no planetary lander has detected Schumann resonance signals. The Huygens probe, which landed on Saturn's largest moon Titan in 2005, detected an unexpected signal at 36 Hz which was initially assumed to be a lightning-induced Schumann resonance, but which was subsequently interpreted as an effect from Saturn's magnetosphere (Beghin et al., 2009). There is no evidence yet for lightning on Titan, see Section 4.2.

*2.3 Acoustic and magnetic signals from lightning*

Acoustic signals from lightning are of course known as thunder, originating from the atmospheric pressure waves set up by the heated lightning channel. The vacuum of space means that sound cannot propagate outside an atmosphere, and therefore thunder is only detectable *in situ*. For *in situ* measurements propagation distance becomes relevant, so thunder can only be detected relatively near the storm. Petculescu and Lueptow (2007) found that acoustic propagation of thunder would be strongest on Titan and weakest on Mars. Huygens carried a microphone (Grard et al., 2006), but it did not detect thunder. Infrasound (<20 Hz) from above-cloud discharges has been detected on Earth (Farges et al, 2005), which suggests a new technique for planetary TLE detection.

Electricity and magnetism are inextricably linked by Ampere's law, so magnetic signals are also emitted by lightning. Normally, electrical signals are preferred for remote sensing, because their amplitude drops off with distance squared rather than distance cubed. However, in the absence of suitable electrical instrumentation, magnetism can be exploited to search for lightning, as was the case on the Venus Express spacecraft, which detected magnetic signals from whistlers (Russell et al., 2007).

## 3. Techniques used to detect planetary lightning

Before discussing the techniques used to detect planetary lightning, it is perhaps worth considering what constitutes a detection of lightning. Whereas Voyager 1's optical and whistler measurements at Jupiter were clearly considered an unambiguous discovery, the situation for some other planets is not so clear-cut. In particular, there has been discussion about the interpretation of microwave and radio emissions from Mars, and lightning on Venus is still hotly debated (see e.g. Yair, 2012). Positive detection of lightning can be provided by electromagnetic evidence backed up by theoretical expectations, for example, optical transients in a predicted spectral region, or radio signals consistent with theoretically-derived propagation characteristics. Simultaneous radio and optical observations would also provide evidence for lightning, perhaps even if one set of measurements were problematic. When unambiguous evidence is not available, the situation becomes more complicated. For example, several sets of electromagnetic signals from the Venusian atmosphere have been linked to lightning, but there are no clear optical signals, despite the rapid inferred flash rate of 18 $s^{-1}$ (Yair, 2012). There are additional theoretical problems; for example, Michael et al (2009) predict that the Venus clouds are too conductive to sustain the charge separation required for lightning. Efforts to improve measurements of Venus lightning are discussed in sections 4 and 5.

Techniques used to detect planetary lightning are summarised in Table 1. The most commonly used technique is a camera on an orbiting satellite. *In situ* measurements offer strong possibilities for unambiguous detection if appropriate instrumentation is included, but there is a large overhead in getting a probe to the planetary surface, so it is a rare occurrence. Huygens carried mutual impedance and relaxation probes, which were intended to measure both lightning and, away from thunderstorms, "fair weather" atmospheric electricity. The antennas in the mutual impedance probe also acted as a radio receiver for any lightning discharges (Grard et al., 2006). An additional consideration for *in situ* measurements is that there is an element of luck in the location and duration of the measurements. For example, the Galileo probe entered an anomalous region of Jupiter's atmosphere and detected neither the anticipated cloud structure, nor optical lightning (Young et al., 1996). However, Galileo's lightning and radio emissions detector (LRD) instrument measured radio signals, presumably from electrical discharges in the planet's atmosphere (Rinnert et al., 1998). Lightning sferics in the VLF range are usually strong radio emissions that can be detected several thousands of kilometres away from the discharge.

Ground-based radio telescopes increasingly have a suitable temporal and spectral response for planetary lightning detection, which can be traded off against the strong signals that would reach a nearby spacecraft, but only during its limited flyby or orbit period. More information is presented on the future prospects for detecting planetary lightning from Earth in section 5. Non-detection from any well-characterised instrument is scientifically valuable, since it can be used to infer quantities such as the maximum lightning flash rate, or the limits of the radio emission spectrum (e.g. Fischer et al., 2007; Gurnett et al., 2010). An indirect detection technique not mentioned in Table 1 is via atmospheric chemical constituents which are assumed to be created by electrical discharges (e.g. Krasnopolsky, 2006).

| **Signals to be detected** | **Earth – based** | **Space – based** | ***In situ*** |
|---|---|---|---|
| Optical | Telescope with filter, often looking at night side | Camera with filter, often looking at night side | Visible transients |
| Other electro-magnetic | Radio telescopes | Radio receivers; Magnetometer | Radio receivers; Magnetometer; Relaxation probe |
| Acoustic | n/a | n/a | Microphone; Sensitive pressure sensor |

**Table 1. Planetary lightning detection technologies**

## 4. Planetary lightning measurements

In this section, we will briefly review the physics behind, and recent observations of, planetary lightning; more see e.g. Yair (2012) and Yair et al. (2008) for more detailed material.

*4.1 Lightning generation mechanisms*

On Earth, there are essentially two mechanisms by which lightning is generated, both of which are anticipated to exist on other planets. Most lightning comes from thunderclouds, which are deep convective systems containing water in several phases. Collisional charge transfer occurs between graupel (soft hail) and ice crystals, and the simplest model is that the ice crystals, usually positively charged, are carried up to the top of the thundercloud and the graupel falls downwards, separating the charges and establishing a potential difference. When this potential difference is large enough, breakdown can occur within the cloud, to the surface, or to a different cloud. The sign of charge transferred is sensitive to the liquid water content in the cloud, both of which can show considerable variability, so there are numerous natural variations in thundercloud charge structure. Additionally, lightning can take a range of forms such as intracloud, intercloud and cloud to ground, with variable electromagnetic characteristics (e.g. Rakov and Uman (2007), MacGorman and Rust (1998)).

The second type of lightning seen on Earth is volcanic lightning, which also arises from convective activity, but in a volcanic plume rather than a cumulonimbus cloud. Volcanic plumes contain large amounts of ash, which can transfer charge by friction (triboelectrification). Smaller particles tend to become negative and larger particles positive, which, like in thunderclouds, can lead to charge separation under gravity and ultimately discharges. Other, smaller scale, processes, such as the charge released by breaking rock (fractoemission) can also cause lightning near the eruption site; see Aplin et al. (2016) for a recent review.

*4.2 Planetary lightning observations*

Table 2 is an up-to-date summary of optical and radio emissions from lightning, which demonstrates that most of the atmospheres in our solar system show some evidence for lightning (also see e.g. Harrison et al., 2008). As discussed earlier, observations of lightning on Venus are marginal, and the "standard" thundercloud charging model may not be relevant, because, although there is convection, the clouds (composed primarily of sulphuric acid) may not contain ice. This does not rule out triboelectric charging of Venus cloud particles, and volcanic lightning may also be possible since Venus seems likely to have had, or even still have, the explosive eruptions that can lead to lightning (Airey et al., 2015). A comprehensive search for short luminous events in the visible on the Venus nightside atmosphere with the Venus Express Visible and Infrared Thermal Imaging Spectrometer (VIRTIS) has found no evidence of lightning (Cardesín Moinelo et al., 2016).

| Planet/Detection | optical | VLF | HF radio |
|---|---|---|---|
| Venus | Yes? | Yes? | Yes? |
| Earth | Yes | Yes | Yes |
| Jupiter | Yes | Yes | No |
| Saturn | Yes | Yes? | Yes |
| Titan (moon) | No | Yes? | No |
| Uranus | No | No | Yes |
| Neptune | No | Yes | Yes? |

**Table 2. Detection of planetary lightning via optical, VLF (very low frequency) radio emissions (whistlers or sferics below ionospheric cut-off frequency) and HF (high frequency) radio emissions above the ionospheric cut-off frequency. Question marks denote ambiguous detections (e.g. small number of signals or other possible explanations).**

Lightning has been observed on multiple occasions at Jupiter and Saturn by a variety of techniques, though it is interesting to note that HF radio emissions have never been seen from Jupiter, due to the properties of its ionosphere (Zarka, 1985), and perhaps also the properties of its lightning (Farrell et al., 1999). Since the lightning on both Jupiter and Saturn is thought to originate from areas of the atmosphere where there are water clouds at similar temperatures to terrestrial clouds, an Earth-like mechanism is assumed. The Cassini spacecraft in orbit around Saturn has monitored lightning activity on Saturn for over a decade. Figure 2 shows the number of Saturn Electrostatic Discharges (SEDs, HF radio emissions detected by the Cassini Radio and Plasma Wave Science (RPWS) instrument) per Saturn rotation as a function of time.

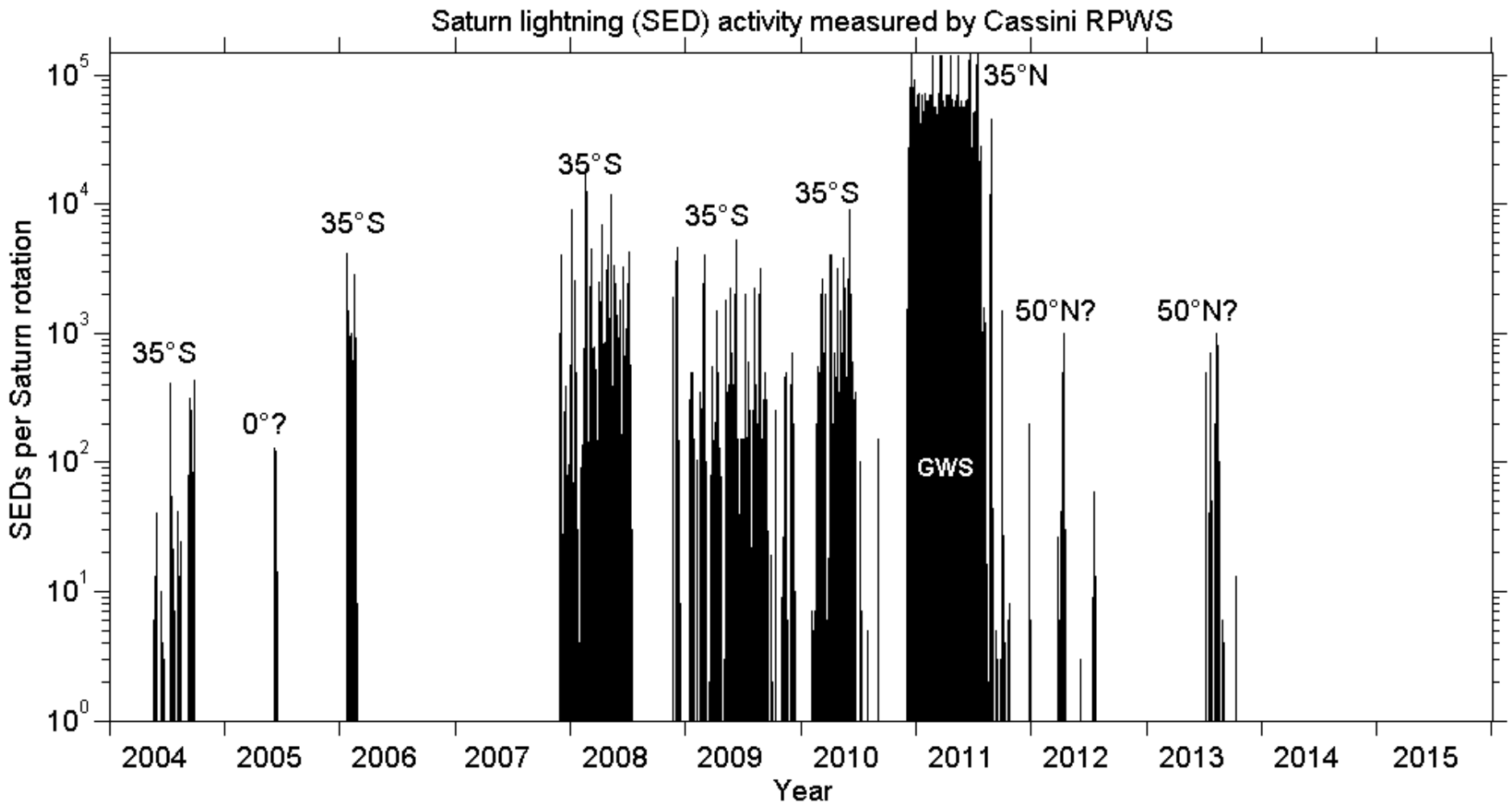

**Figure 2. Detected number of Saturn lightning radio bursts (per Saturn rotation of 10.66 h) per year. Lightning storms have mostly been located at 35°S and possibly also at the equator and 50°N. The so-called "Great White Spot" (GWS) was located at 35°N.**

Figure 2 shows that Saturn lightning storms can last from a few days up to several months, with the longest lightning storm lasting for 11 months, almost all of 2009. Their occurrence could be related to Saturn's seasons as most of the storms appeared ±2 years around Saturn equinox (August 2009), with a switch in location from the southern to the northern hemisphere. However, since sunlight only penetrates down to the 1-2 bar level and the flashes are thought to originate from a depth of 8-10 bars (Dyudina et al., 2010; 2013), sunlight might only act as a trigger for instability and internal convection, with the storms powered by Saturn's internal heat. There are two different classes of Saturn lightning storms: the smaller storms have a horizontal extension around 2000 km and a flash rate of $10^3$-$10^4$ per Saturn rotation, and there are the rare and giant "Great White Spots" (GWS) that usually occur only once per Saturn year (29.5 Earth years) with a flash rate that is about an order of magnitude higher than the smaller storms. The GWS of 2010/2011 (see Figure 3) had a latitudinal extension of 10,000 km, and its elongated eastward tail was wrapped all around the planet, which is a distance of 300,000 km (Fischer et al., 2011).

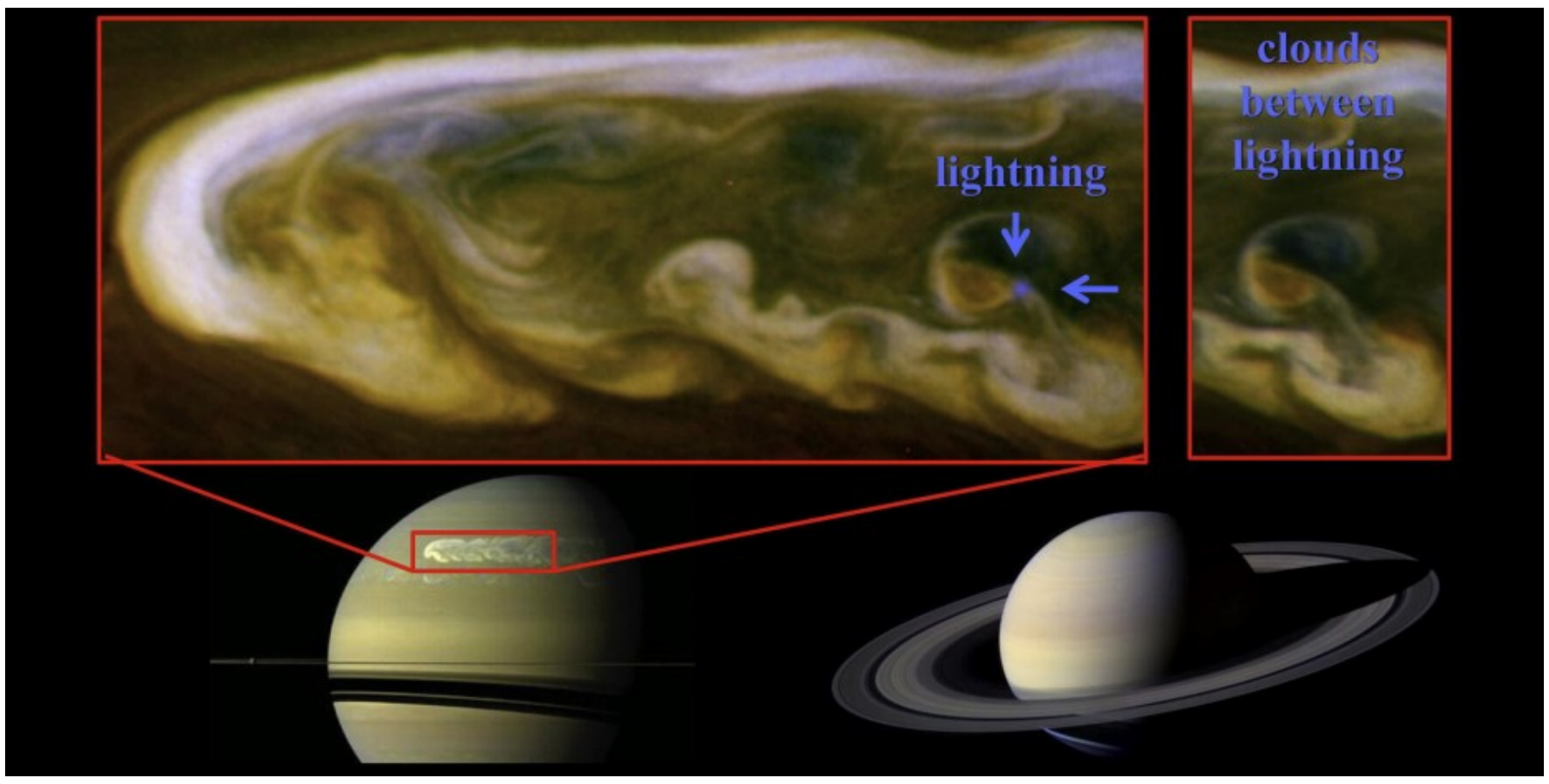

**Figure 3. Optical detection of Saturn lightning on the dayside in the tail of the Great White Spot (Reproduced with permission from Dyudina et al., 2013).**

Saturn lightning flashes were first detected optically on the night side around equinox, when reflected light from the rings towards the planet was minimised. The Cassini cameras spotted flash-illuminated cloud tops with a diameter of ~200 km, suggesting that the lightning is 125-250 km below (Dyudina et al., 2010). The optical energy of a single Saturn flash is about $10^9$ J, comparable to the optical energy of a Jovian flash. Later, more sophisticated image processing revealed visible lightning on the dayside in blue light in the tail of the GWS as illustrated in Figure 3 (Dyudina et al., 2013).

The possibility of lightning in the nitrogen-methane atmosphere of Titan, Saturn's largest satellite, has been studied for many years, e.g. Tokano et al. (2001). However, the search for HF radio signals performed by Cassini RPWS during the Titan flybys (Fischer et al., 2007) brought no positive results after more than 100 flybys. There are unidentified impulsive signals in the frequency range up to 11.5 kHz in the data of the Atmospheric Structure Instrument recorded by the Huygens Probe during its descent onto Titan in January 2005, which could be due to lightning (Fulchignoni et al., 2005). Since there are rare outbursts of convective methane clouds in Titan's atmosphere, Cassini will continue its search for Titan lightning until the final, 126$^{th}$, Titan flyby.

The planets beyond Saturn have only been visited by one spacecraft, Voyager 2, which encountered Uranus in 1986 and Neptune in 1989. Although not designed specifically for this purpose, Voyager 2's Planetary Radio Astronomy (PRA) and Plasma Wave Instruments were able to respond to a range of radio emissions from lightning. Specifically, so-called Uranian Electrostatic Discharges (UEDs), bursts of broadband radio emission, were detected by the PRA instrument (Zarka and Pedersen, 1986). Detection events from Neptune lightning were more marginal, with only a few events detected as whistlers and at higher frequency by the plasma wave instrument. There was no optical detection at either planet, but other atmospheric observations, and modelling, point towards mixed-phase water clouds in both atmospheres, which could support lightning generation by a similar mechanism to Earth (Yair et al., 2008).

Mars is a notable omission from Table 2 since lightning has not yet been observed there, though it is expected to exist, and the planet is also a prime contender for a global electric circuit if lightning is detected. The dusty, windy carbon dioxide atmosphere is known to harbour dust devils, rotating convectively-generated dust spirals. On Earth, dust devils generate substantial electric fields of ~100 $kVm^{-1}$, but as this does not approach the breakdown voltage of air (3 $MVm^{-1}$) no lightning is seen in them. On Mars however, the atmosphere is lower pressure, which reduces the breakdown voltage to 20-30 $kVm^{-1}$, easily reached by terrestrial dust devils. Discharges within dust devils or large-scale dust storms would be difficult to detect optically, and radio emission searches have not yielded a positive signal (Gurnett et al., 2010). So far, no relevant *in situ* instrumentation has been sent to Mars, but this should change – at the time of writing, the ExoMars mission is on its way to Mars.

## 5. Future prospects for planetary lightning detection

ExoMars was launched on 14 March 2016, carrying an orbiter and a lander. The lander carries a suite of environmental sensors, DREAMS (Dust characterization, Risk Assessment, and Environment Analyser on the Martian Surface), which includes MicroARES (Atmospheric Radiation and Electricity Sensor). MicroARES is a relaxation probe that can measure atmospheric conductivity and both AC and DC electric fields, and will provide the first *in situ* electrical measurements (Harrison et al., 2016). Measurements are planned to take place over a few days in October 2016, during dust storm season.

The Japanese Akatsuki mission to Venus was launched in 2010, but failed to reach orbit immediately. Orbit insertion was successful in December 2015 and at the time of writing, the instruments were being prepared for deployment. The mission aims to understand the dynamics and cloud formation of the Venus atmosphere with a suite of optical instruments. On the spacecraft is a lightning and airglow camera (LAC), which will make fast (32 kHz) observations with the 777.4 nm filter on the night side. The instrument is sensitive enough to detect lightning one hundredth as bright as terrestrial lightning from 1000 km (Nakamura et al., 2011).

Lightning detection events from Akatsuki can be further substantiated by observations with large ground-based radio telescopes. For Saturn, the very intense SEDs (about $10^4$ times stronger than radio emissions from terrestrial lightning at a few MHz) were first recorded from Earth with the large Ukrainian radio telescope UTR-2 (Konovalenko et al., 2013). The spectral flux density of SEDs at Earth is a few hundred Jansky (1 Jy=$10^{-26}$ W $m^{-2}$ $Hz^{-1}$), and the typical sensitivity of UTR-2 is a few Jy, depending on the frequency bandwidth and the integration time of the radio receiver. Assuming that Venus and Earth lightning have similar radiated energies of 0.01 W/Hz at frequencies of a few MHz (Gurnett et al., 1991), the flux density of Venus lightning at Earth when the planets are closest would be 45 Jy. This is detectable by UTR-2, and the radio flux scales linearly with radiated energy. Similarly, the radio flux of UEDs from Uranus at Earth should be a few Jy, close to the sensitivity of UTR-2. Therefore, the planets Venus and Uranus are the most promising candidates for future detection of HF radio emissions from lightning with large ground-based radio telescopes. Ground-based measurements will permit observations of Saturn lightning even after the end of the Cassini mission in September 2017. Radio telescopes can measure the SEDs, and amateur astronomers are also able to spot the prominent 2000 km-sized thunderstorms on Saturn with their backyard optical telescopes. Ground-based observations of the dynamical atmospheres of the gas planets will become increasingly important in the next decade since the next mission to an outer planet, the European JUICE (Jupiter Icy Moon Explorer) spacecraft, is planned to arrive at Jupiter in 2030.

## 6. Conclusions

Planetary lightning detection technology has developed considerably since the first observations at Jupiter in the late 1970s. Space instruments for lightning detection have focused on radio and optical detection techniques. Lightning information has also been opportunistically exploited from plasma and magnetism instruments, to extract the maximum scientific return from each mission.

Despite the many lightning observations made at Jupiter and Saturn there is still little understanding of charge separation mechanisms in planetary atmospheres, with heavy reliance on terrestrial analogies. Measurements anticipated at Venus and Mars in 2016 are likely to increase our understanding of electrical processes in those environments. Looking further into the future, ground-based detection offers prospects for longer-term measurements of planetary lightning, particularly from Venus and the outer planets.

**Acknowledgement:** G.F. was supported by a grant (P24325-N16) from the Austrian Science Fund FWF. The authors have no conflicts of interest.